\documentclass[letterpaper]{jpconf}
\usepackage{graphicx}
\begin{document}

\title{Searching for High Energy Diffuse Astrophysical Muon Neutrinos with IceCube}

\author{Sean Grullon for the IceCube Collaboration}

\address{University of Wisconsin - Madison}

\begin{abstract}
The IceCube Neutrino Observatory is a 1 $km^{3}$ detector currently 
under construction at the South Pole. Searching for high energy 
neutrinos from unresolved astrophysical sources is one of the main 
analysis strategies used in the search for astrophysical neutrinos with 
the IceCube Neutrino Observatory.  A hard energy spectrum of 
neutrinos from isotropically distributed astrophysical sources could 
contribute to form a detectable signal above the atmospheric 
neutrino background.  A reliable method of estimating the energy of 
the neutrino-induced lepton is crucial for identifying astrophysical 
neutrinos. An analysis is underway using data from the half completed detector  
conÞguration taken during its 2008-2009 science run. 
\end{abstract}

\section{Neutrinos As Cosmic Messengers}

	There are many objects in our universe that involve extremely high energy processes. These processes involve the accretion of matter into black holes at the centres of active galaxies, supernovae, and gamma-ray bursts, where an enormous amount of energy is released over time scales as short as a few seconds.  Understanding these astrophysical objects and the underlying physics involves observing high energy radiation.  The three particle messengers involved in high-energy astronomy are charged cosmic rays (protons and nuclei), gamma-rays, and neutrinos.  Cosmic ray and gamma-ray astrophysics have been extremely successful fields, however the nature of their sources is still not completely understood.  Neutrinos may elucidate the fundamental connection between the sources of high energy cosmic rays and gamma-rays. 

	Cosmic rays - high energy protons and nuclei - have been well studied with both space and ground based instruments.  Their major astronomical disadvantage is that they are charged particles and thus are deflected by magnetic fields subsequently losing their directionality.  High energy gamma-rays have been detected for many galactic and extra galactic objects, but their effectiveness as cosmic messengers over long distance scales is limited by their absorption by extra-galactic background light. Neutrinos could provide a fundamental connection between cosmic rays and gamma-rays. If a gamma-ray source was found to be a neutrino source, then a hadronic accelerator central engine might be simultaneously driving cosmic ray, gamma-ray, and neutrino production in one astrophysical object \cite{learnedmannheim:2000}.
	
	A next generation kilometer scale neutrino observatory, IceCube, \cite{performancepaper:2006} is currently under construction at the geographic South Pole.  When construction is completed in January 2011, IceCube will consist of an in-ice cubic kilometer neutrino detector as well as a square kilometer cosmic ray air shower array at the surface called IceTop.  The in-ice detector consists of photomultiplier tubes deployed in an array of strings.   IceTop consists of an array of stations with two tanks filled with ice and four phototubes at every station.  Construction began at the South Pole during the austral summer 2004-05, with 1 in-ice string and 4 IceTop stations deployed  \cite{performancepaper:2006}. Finally, IceCube will consist of 86 strings giving a total of 5160 phototubes and 80 IceTop stations with 4 phototubes per station.   
	
	The backgrounds to search for a flux of high-energy astrophysical neutrinos are atmospheric muons and neutrinos from the interaction of cosmic rays in the Earth's atmosphere  \cite{honda:2006}.  Atmospheric muons can be eliminated by looking for events moving upward through the detector.  A fraction of the downgoing muon flux will be falsely reconstructed in the upward direction, but can be removed by tight requirements on the fitted track.  An event selection based on these tight requirements results in a rather pure sample of atmospheric neutrinos.  This flux of atmospheric neutrinos seen in the IceCube array is the main background to an astrophysical search.  
		
\section{Astrophysical Neutrino Point Source Search}

	Searching for resolved sources of astrophysical muon neutrinos is one of the main science goals of the IceCube neutrino observatory.  A search for point sources of neutrinos is made by looking for an excess of events from a specific direction in the sky.  After elimination of background atmospheric muons by using the Earth as a filter and making a high energy cut, the skymap of the arrival directions of the neutrino-induced muons is analyzed for evidence of astrophysical neutrinos.  Astrophysical neutrinos are expected to show up as an excess over background which is found by looking at the average rate of events in the same declination band.  A significance of the observation is found by comparing the number of events in the on-source bin to that predicted from the off-source band.  A full maximum likelihood method is used to search for an excess.  The method uses the measured event by event directional likelihood resolution as a parameter in the likelihood fit.  The reconstructed event energy is also used in the fit to allow to determine the slope of the energy spectrum of a neutrino source.  

\begin{figure}[htp]
\centering
\includegraphics[scale=0.22]{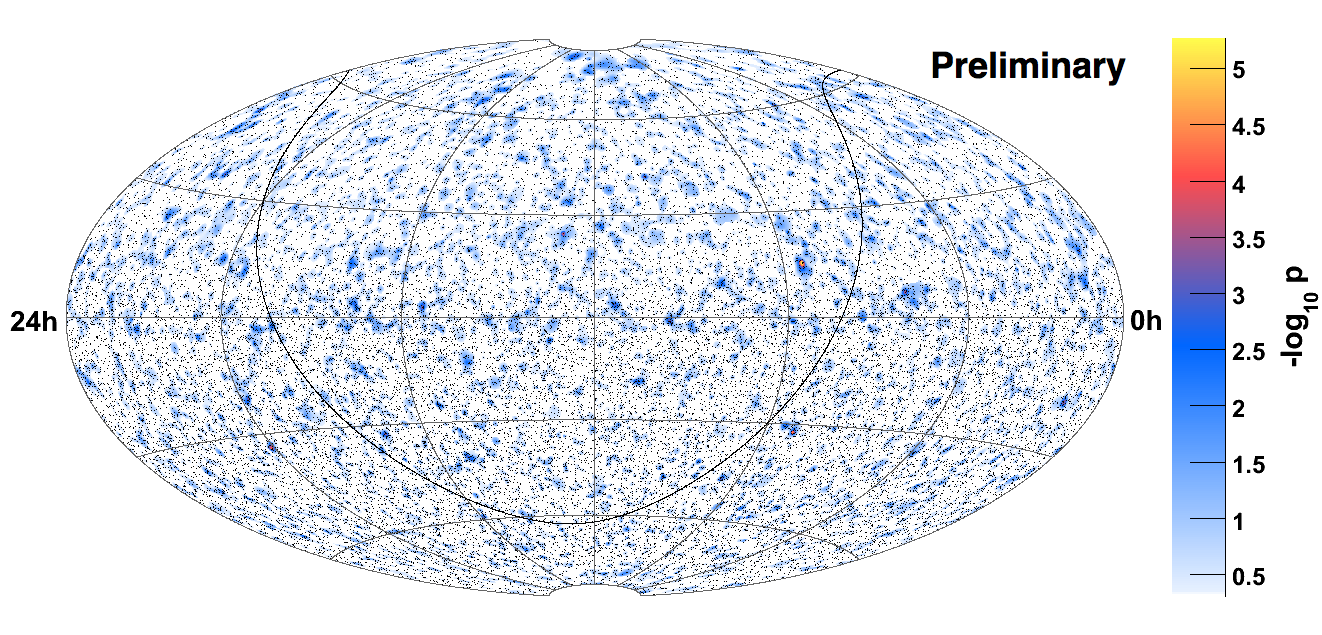}
\caption{Skymap for the IceCube detector in the 40 string configuration for one year of data taken during 2008. }\label{skymap}
\end{figure}

	Several types of searches are performed \cite{dumm:2009}. The most general is an all-sky search where every point across a fine grid in the sky is taken as a possible source position.  The likelihood is used to find the best fit to the number of astrophysical neutrinos and the spectral shape.  Other searches include looking for an excess amongst a pre-specified list of possible sources, or to stack a class of objects to see if the superposition of these objects gives significant excess.   The data from the half completed detector (40 string) was analyzed to produce a sky map covering the full sky range from 0 to 180 degrees in zenith.  After analysis, no evidence for sources was seen as shown in Fig. \ref{skymap}.  The most significant spot in the sky had a significance consistent with a random fluctuation. 
		
\section{Diffuse Astrophysical Neutrino Flux Search}

If individual sources of astrophysical neutrinos are unresolved, then we may still perform a search for astrophysical neutrinos by looking for an excess of events over the whole sky above the expected atmospheric neutrino background rate.  A superposition of all neutrino sources in the universe would give rise to an extra-terrestrial flux that has a harder energy spectrum than that of atmospheric neutrinos.  Since predictions for the astrophysical flux go as $dN/dE\sim E^{-2}$ \cite{learnedmannheim:2000}, one looks for higher energy events in the detector to separate them from the steeper atmospheric neutrino spectrum that goes as $dN/dE\sim E^{-3.7}$  \cite{honda:2006}. In contrast to the point source search, the diffuse search requires a detailed understanding of the atmospheric neutrino background and detector sensitivity through Monte Carlo simulation in order to correctly interpret the observed neutrino events.  

\subsection{Event Energy Reconstruction}

Since extra-terrestrial source of neutrinos are expected to have harder energy spectra than the atmospheric neutrino backgrounds, a reliable method for reconstructing the energy of the event is crucial.  Historically this measure was a simple one - counting the number of array photomultiplier tubes that detected light.  This was a reasonably powerful energy estimator, however research continued on better algorithms to improve the correlation between true and reconstructed energy \cite{grulloninternalreport:2008}. The reconstruction algorithm used here models a muon with constant energy loss per unit length.  The energy is adjusted to minimize the log-likelihod of the observed phototube charge densities given the expectation from the model.  
	 
\subsection{Analysis Method}

\begin{figure}[htp]
\includegraphics[scale=0.25]{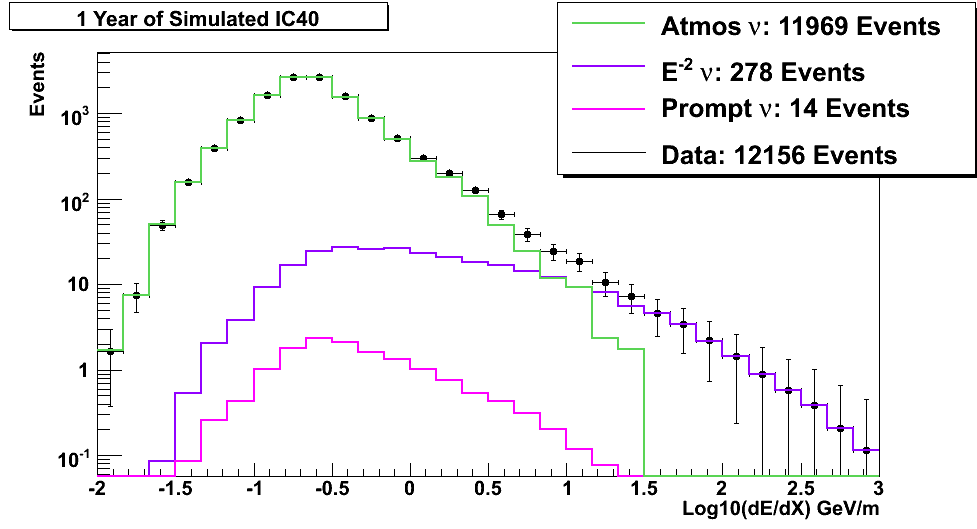}
\includegraphics[scale=0.25]{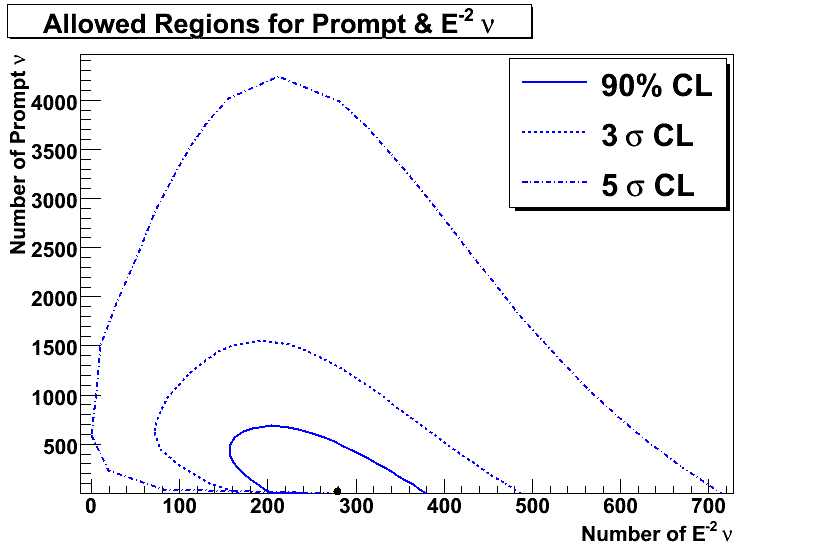}
\caption{Methodology of a diffuse search, demonstrated with a random simulated data set drawn from respective simulation samples.  The observed reconstructed energy distribution (left) is interpreted as allowed regions of the number of astrophysical and prompt neutrinos (right) using a parametrized maximum likelihood fit.} \label{fitexample}
\end{figure}

After data quality cuts remove poorly reconstructed downgoing muons, the reconstructed energy distribution of the candidate neutrino events in the detector is analyzed.  A likelihood framework fits contributions from conventional atmospheric neutrinos (resulting from pion and kaon decays), prompt atmospheric neutrinos (resulting from charmed meson decays) and an astrophysical signal flux to the data.  The systematic uncertainties of the detector are treated as nuisance parameters in the likelihood fitting procedure.  These nuisance parameters are allowed to float within their understood error ranges in the fit.  The resulting confidence regions of the physics parameters are then examined to see if the fit favors a background only hypothesis or a signal hypothesis.  An example is given in Fig. \ref{fitexample}.  Here, a random sample of events is drawn according to Poisson probabilities from the simulated data distribution.  This sample is then analyzed as if it were real data.  The confidence regions for the contributions of prompt and astrophysical neutrinos are also shown in Fig. \ref{fitexample}. These confidence regions show the discovery potential of IceCube in the 40 string configuration for the case of two physics parameters. These two parameters are the number of prompt atmospheric neutrinos and the number of astrophysical neutrinos, respectively.  

\begin{figure}[hbp]
\centering
\includegraphics[scale=0.375]{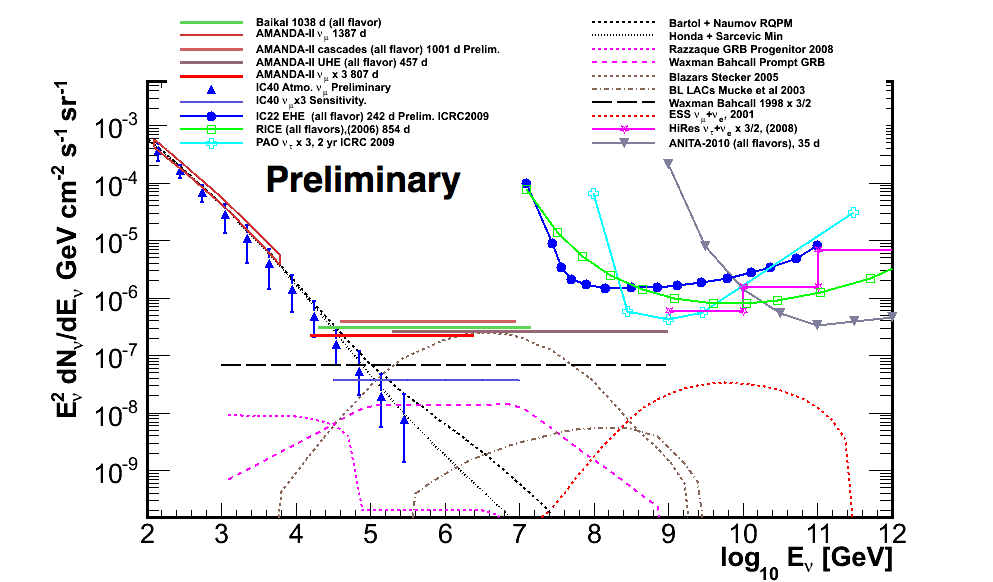}
\caption{Diffuse neutrino fluxes and experimental limits. The sensitivity of IC40 is just below the Waxman-Bahcall bound. References for the curves are given in \cite{becker:2007}. } \label{limits}
\end{figure}

At this time, to avoid possible biases in the analysis, the highest energy events in the dataset are kept hidden as the method is being developed and the sources of systematic uncertainty understood.  Characterizing how the response of the detector is affected by the optical properties of the South Pole ice is especially important for the systematics.  The sensitivity of this analysis for the 40 string detector is shown in Fig. \ref{limits}.

\section{Outlook} 
The construction of a kilometer scale, high energy neutrino telescope is nearing completion at the South Pole.   IceCube will hopefully lead to new discoveries about the nature of the high energy universe. Searches for point sources and diffuse fluxes of astrophysical neutrinos will either result in some these discoveries or set very stringent limits on the underlying physics of high energy neutrinos in the cosmos.        
     
\section{References}
\bibliographystyle{iopart-num}
\bibliography{LLWIproceedingsGrullon}

\providecommand{\newblock}{}
\begin{thebibliography}{1}
\expandafter\ifx\csname url\endcsname\relax
  \def\url#1{{\tt #1}}\fi
\expandafter\ifx\csname urlprefix\endcsname\relax\def\urlprefix{URL }\fi
\providecommand{\eprint}[2][]{\url{#2}}
% Bibliography created with iopart-num v2.0
% /biblio/bibtex/contrib/iopart-num

\bibitem{learnedmannheim:2000}
Learned J and Mannheim K 2000 {\em Ann. Rev. Nucl. Part. Science\/} {\bf 50}
  679

\bibitem{performancepaper:2006}
Achterberg A {\em et~al.\/} 2006 {\em Astropart. Physics\/} {\bf 26} 155

\bibitem{honda:2006}
Honda M {\em et~al.\/} 2007 {\em Phys. Rev. D\/} {\bf 75} 043006

\bibitem{dumm:2009}
Dumm J 2009 {\em Proceedings of the 31st International Cosmic Ray Conference,
  Lodz, Poland\/}

\bibitem{grulloninternalreport:2008}
S~Grullon D J Boersma G~Hilll K~H and Mase K 2007 {\em Proceedings of the 30th
  International Cosmic Ray Conference, Merida, Mexico\/} p 159

\bibitem{becker:2007}
Becker J 2007 {\em Neutrinos on the rocks: On the phenomenology of potential
  astrophysical neutrino sources\/} Ph.D. thesis Dortmund University of
  Technology

\end{thebibliography}

\end{document}